\documentclass[prb,aps,superscriptaddress,reprint
]{revtex4-2}
\usepackage{siunitx}
\usepackage{graphicx}
\usepackage{amsmath}
\usepackage[english]{babel}
\usepackage[colorlinks=true,urlcolor=blue,linkcolor=blue,citecolor=blue]{hyperref}
\usepackage{mathptmx}
\makeatletter
\DeclareRobustCommand{\equalcontrib}{%
  \frontmatter@footnote{These authors contributed equally to this work.}%
}
\makeatother

\begin{document}

\title{Extraordinary high room-temperature carrier mobility \\ in graphene-WSe$_2$ heterostructures}

%%%%%%%%%%%%%%%%%%%%%%%%

\author{Luca Banszerus\equalcontrib}
\thanks{Present address: Faculty of Physics, University of Vienna}
\affiliation{JARA-FIT and 2nd Institute of Physics, RWTH Aachen University, 52074 Aachen, Germany}

\author{Taoufiq Ouaj\equalcontrib}
\affiliation{JARA-FIT and 2nd Institute of Physics, RWTH Aachen University, 52074 Aachen, Germany}

\author{Thibault Sohier}
\affiliation{Laboratoire Charles Coulomb (L2C), Université de Montpellier, CNRS, Montpellier, France}

\author{Alexander Epping}
\affiliation{JARA-FIT and 2nd Institute of Physics, RWTH Aachen University, 52074 Aachen, Germany}

\author{Florian Winkler}
\affiliation{Ernst Ruska-Centre for Microscopy and Spectroscopy with Electrons,  Forschungszentrum Jülich, 52425 Jülich, Germany}

\author{Mrityunjay Pandey}
\affiliation{JARA-FIT and 2nd Institute of Physics, RWTH Aachen University, 52074 Aachen, Germany}

\author{Frank Volmer}
\affiliation{JARA-FIT and 2nd Institute of Physics, RWTH Aachen University, 52074 Aachen, Germany}

\author{Karl Leuckefeld}
\affiliation{JARA-FIT and 2nd Institute of Physics, RWTH Aachen University, 52074 Aachen, Germany}

\author{Florian Libisch}
\affiliation{Institute for Theoretical Physics, Vienna University of Technology, 1040 Vienna, Austria}

\author{Federica Haupt}
\affiliation{Advanced Microelectronic Center Aachen, AMO GmbH, 52074 Aachen, Germany}

\author{Marvin Metzelaars}
\affiliation{Institute of Inorganic Chemistry, RWTH Aachen University, 52056, Aachen, Germany}

\author{Paul Kögerler} 
\affiliation{Institute of Inorganic Chemistry, RWTH Aachen University, 52056, Aachen, Germany}
\affiliation{Peter Grünberg Institute, Electronic Properties (PGI-6) Forschungszentrum Jülich, 52425, Jülich, Germany}

\author{Kenji Watanabe}
\affiliation{Research Center for Electronic and Optical Materials, National Institute for Materials Science, 1-1 Namiki, Tsukuba 305-0044, Japan}

\author{Takashi~Taniguchi}
\affiliation{Research Center for Materials Nanoarchitectonics, National Institute for Materials Science,  1-1 Namiki, Tsukuba 305-0044, Japan}

\author{Knut Müller-Caspary}
\affiliation{Ernst Ruska-Centre for Microscopy and Spectroscopy with Electrons,  Forschungszentrum Jülich, 52425 Jülich, Germany}

\author{Nicola Marzari}
\affiliation{Theory and Simulation of Materials (THEOS), and National Centre for Computational Design and Discovery of Novel Materials (MARVEL), École Polytechnique Fédérale de Lausanne, 1015 Lausanne, Switzerland}

\author{Francesco Mauri}
\affiliation{Dipartimento di Fisica, Universita di Roma La Sapienza, 00185 Roma, Italy  and Graphene Labs, Fondazione Istituto Italiano di Tecnologia, 16163 Genova, Italy}

\author{Bernd Beschoten} 
\affiliation{JARA-FIT and 2nd Institute of Physics, RWTH Aachen University, 52074 Aachen, Germany}

\author{Christoph Stampfer}
\altaffiliation{To whom correspondence should be addressed; \\E-mail: stampfer@physik.rwth-aachen.de.}
\affiliation{JARA-FIT and 2nd Institute of Physics, RWTH Aachen University, 52074 Aachen, Germany}
\affiliation{Peter Gr\"unberg Institute (PGI-9) Forschungszentrum J\"ulich, 52425 J\"ulich, Germany}

\begin{abstract}

High charge carrier mobilities play a fundamental role for high-frequency electronics, integrated optoelectronics as well as for sensor and spintronic applications, where device performance is directly linked to the magnitude of the carrier mobility. 
Van der Waals heterostructures formed by graphene and hexagonal boron nitride (hBN) already outperform all known materials in terms of room temperature mobility.
Here, we show that the room temperature carrier mobility of today's best graphene/hBN devices can be surpassed by more than a factor of three by heterostructures formed by tungsten diselenide (WSe$_2$), graphene and hBN, which can have mobilities as high as 350,000~cm$^2$/(Vs) and resistivities as low as $15$~Ohm.
The resistivity of these devices shows a significantly weaker temperature dependence than the one of graphene on any other known substrate.
Notably, the reduced temperature dependence and the resulting mobility enhancement in graphene show an unexpected relation to the thickness of the WSe$_2$ layer and to the minimum conductivity at the charge neutrality point, suggesting a role of increased charge disorder and/or enhanced screening and questioning the current understanding of electron–phonon scattering in graphene-based van der Waals heterostructures.
\end{abstract}

\maketitle

\begin{figure*}[t]
    \begin{center}
\includegraphics{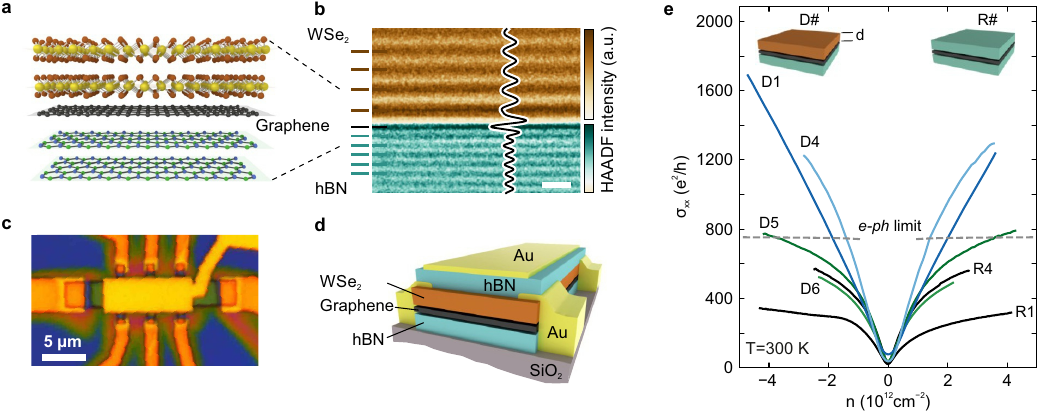}
    \end{center}
    \caption{\textbf{Room-temperature transport in WSe$_2$/Gr/hBN.}
    \textbf{a} Schematic illustration of a WSe$_2$/Gr/hBN heterostructure. 
    \textbf{b} False-color cross-sectional scanning transmission electron microscopy image of one of our samples (D1) highlighting the excellent interface quality and the layer spacing (see line profile plotted as inset). The scale bar is 1~nm.
    \textbf{c} Optical-microscope image of a dual-gated WSe$_2$/Gr/hBN multi-terminal Hall-bar device with metallic top gate (device D1). 
    \textbf{d} Schematic cross-section of the device.
    \textbf{e} Room-temperature longitudinal conductivity of WSe$_2$/Gr/hBN devices (D1, D4, D5 and D6 with WSe$_2$ of different thicknesses $d$, see table~S2 in the SI for details) and of hBN/Gr/hBN reference samples (R1, R4) as a function of charge carrier density $n$.
    Note that devices D1 and D4 exhibit conductivity values that significantly exceed what is commonly considered the upper limit imposed by electron–phonon ({\it e-ph}) scattering on the room-temperature conductivity of graphene (grey dashed line).
    }
    \label{fig:fig_1}
\end{figure*}
%Introduction

Stacking two-dimensional materials into van der Waals heterostructures enables the engineering of
their electronic, magnetic and optical properties~\cite{Geim2013Jul}.
For example, the interaction between graphene (Gr) and hexagonal boron nitride (hBN) can modify graphene’s electronic band structure~\cite{Dean2013May,Hunt2013Jun}, while twisting individual sheets in bilayer graphene can result in correlated insulating states~\cite{Cao2018Amar} or in superconductivity~\cite{Cao2018Mar} and allows to tune many-body interactions. 
Stacking graphene on different materials, such as, for example, transition metal dichalcogenides, can lead to significant changes in nanometre-scale strain variations~\cite{Banszerus2017Feb}, in the phonon
dispersion relation~\cite{Lui2015Apr, Narang2017Aug, Greener2018Aug}, and in electron-electron and electron-phonon coupling in such heterostructures~\cite{hwang2012,faugeras2015,ponomarenko2024,veyrat2020,xin2023,Forster2013Aug}.
All of this significantly impacts carrier mobility in graphene, and a better understanding of their role has led to higher carrier mobilities.

The performance of early graphene devices on silicon dioxide (SiO$_2$) was limited by surface roughness, trapped charges, and remote interface phonon scattering. This resulted in carrier mobilities at room temperature typically falling below 10,000~cm$^2$/(Vs)~\cite{Chen2008Mar}.
Introducing hBN as both a substrate and an encapsulation material overcame many of these limitations, enabling graphene devices to reach room-temperature mobilities of up to 100,000~cm$^2$/(Vs)~\cite{Dean2010Aug,Wang2013Nov}.
Such hBN/Gr heterostructures are now regarded as state-of-the-art, with performance commonly assumed to be limited by scattering with in-plane acoustic and optical phonons~\cite{Sohier2014Sep}.
However, recent theoretical and experimental work has revealed that the coupling to intrinsic optical phonons has been significantly underestimated~\cite{venanzi2023,graziotto2024, Guandalini2025Apr}.
Since the surrounding dielectric environment influences optical phonon scattering, this suggests a stronger influence of the substrate than was previously assumed, and highlights its role in electron–phonon interactions. Therefore, the ultimate limit of graphene's room temperature mobility still remains unclear.

In this work, we 
report the observation that WSe$_2$/Gr/hBN heterostructures 
exhibit room-temperature mobilities that are significantly higher than the upper limit predicted by current theory for pristine graphene~\cite{hwang2012,Sohier2014Sep}.
Interestingly, we find that the thickness of the WSe$_2$ flakes used plays a crucial role in the observed mobility enhancement in these WSe$_2$/Gr/hBN devices.
Devices based on WSe$_2$ flakes thicker than 20 nm exhibit an unusually weak temperature dependence and reach resistivities as low as 15~Ohm, indicating a substantial modification of the underlying electron-phonon scattering processes.

The devices were fabricated using both exfoliated and chemical vapor deposition (CVD)-grown graphene, encapsulated between hBN and WSe$_2$ crystals of varying thickness (see schematic in Fig.~\ref{fig:fig_1}a). 
For the CVD graphene devices (D1-D3), graphene was delaminated from its copper substrate using a 30 to 50~nm thick WSe$_2$-flake~\cite{Banszerus2015Jul,Banszerus2017Feb}. 
The exfoliated graphene devices (D4-D8) were assembled by picking up graphene with pre-stacked hBN/WSe$_2$ half-stacks, where the thickness of WSe$_2$ ranged from 3 to 40~nm.
All obtained heterostructures are deposited onto an hBN-flake placed on SiO$_2$ on a highly-doped silicon substrate~\cite{Banszerus2015Jul,Banszerus2017Feb}, which acts as back gate.
Cross-sectional scanning transmission electron microscopy (STEM)  measurements reveal clean interfaces between both hBN/Gr and Gr/WSe$_2$, as indicated in Fig.~\ref{fig:fig_1}b. 
The layer spacing between graphene and the first WSe$_2$ layer is $5.3 \pm 0.3$~\AA\, (see Supplementary Methods), in good agreement with {\em ab-initio} calculations but substantially smaller (by $\approx$1.5~\AA) than previously reported for bulk WSe$_2$ on hBN~\cite{Rooney2017Sep}.

The heterostructures were shaped into Hall bars with one-dimensional Cr/Au side contacts. 
For comparison, we also fabricated graphene reference devices based on regular hBN/Gr/hBN heterostructures (R1, R2, R4-R6)~\cite{ouaj2023,ouaj2024}.
One of the WSe$_2$/Gr/hBN devices features an additional hBN-layer and an additional metallic top gate (see  Fig.~\ref{fig:fig_1}c,d), allowing for an independent control of the carrier density in WSe$_2$ and graphene, respectively. 
All electrical transport measurements were performed using lock-in techniques in a four-probe geometry (see Ref.~\cite{PhysRevApplied.18.014028} for details). 

The double-gated device D1 enables us to distinguish the graphene contribution from possible parallel conduction through WSe$_2$.  We find that, away from graphene's charge neutrality point, WSe$_2$ does not contribute appreciably to the measured conductivity, apart from a weak back-gate-independent parallel channel (Supplementary Methods). 
This is further confirmed by quantum Hall measurements and Hall-effect characterization, which show that transport in devices D1-D4 is dominated by graphene. Parallel conduction through WSe$_2$ can therefore be excluded as the origin of the observed behavior.

The room-temperature longitudinal conductivity $\sigma_\text{xx}$ of six representative devices is shown in Fig.~\ref{fig:fig_1}e as a function of graphene's charge-carrier density $n$,
which has been estimated by the electrostatic coupling to the back gate  using both Hall and quantum Hall measurements (see Supplementary Methods).
The conductivities of the reference
hBN/Gr/hBN samples (R1 and R4; black traces) exhibit the typical behaviour of state-of-the-art high-mobility graphene devices, i.e., it saturates at conductivity values of a few hundred e$^2$/h at large carrier densities, consistent with scattering being dominated and limited by electron-phonon scattering~\cite{Dean2010Aug,Wang2013Nov}.
By contrast, the conductivity of the WSe$_2$/Gr/hBN device D1 (based on CVD graphene and an 31~nm thick WSe$_2$ flake; blue trace) shows %a symmetric 
an almost linear behavior over the full electron ($n>0$) and hole ($n<0$) regime, reaching values more than four times higher than those of the reference samples, corresponding to resistivity values as low as 15~Ohm (1600~$\mathrm{e^2/h}$). 
A similarly high conductivity is observed for the WSe$_2$/Gr/hBN device D4 (light blue trace), which is based on exfoliated graphene and a WSe$_2$ flake with a thickness of around 40~nm. This indicates that the enhanced conductivity is not limited to a specific fabrication method.
Rather, it appears that the thickness of the WSe$_2$ layer plays a more decisive role: the WSe$_2$/Gr/hBN devices D5 and D6, which incorporate comparatively thin WSe$_2$ flakes (3 and 15~nm; green traces), exhibit significantly reduced conductivity at higher carrier densities.

\begin{figure}[t]
    \begin{center}
    \includegraphics{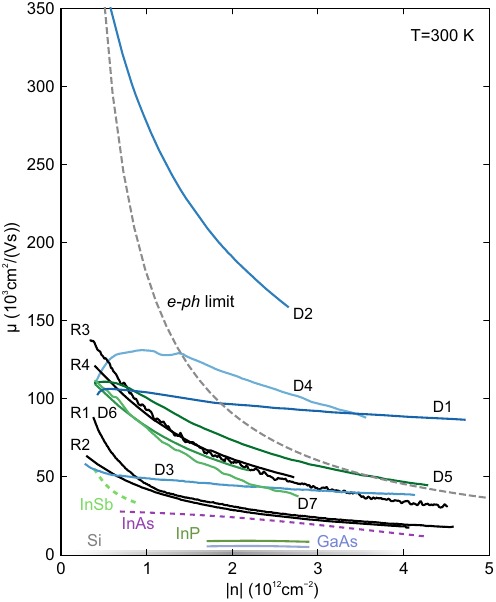}
    \end{center}
    \caption{\textbf{Carrier mobility as a function of charge carrier density} of seven devices based on WSe$_2$/Gr/hBN (blue traces: thick WSe$_2$, green traces: thin WSe$_2$), and four reference devices (R1-R4) based on hBN/Gr/hBN (black traces), including data from Ref.~\cite{Wang2013Nov} (labelled as R3). The gray-dashed line indicates our current understanding of the mobility limit set by electron-phonon ({\it e-ph}) scattering. This limit is well established both in experiment~\cite{Hwang2008Mar,Chen2008Mar,Dean2010Aug,Zou2010Sep,Efetov2010Dec,Wang2013Nov} and theory~\cite{Pietronero1980Jul,Hwang2008Mar,Perebeinos2010May,Park2014Mar,Sohier2014Sep}, with uncertainties of the order of $20\%$ at most. The remaining traces indicate the range of carrier mobilities reported in the literature~\cite{Wang2013Nov} for high-performance III-V semiconductors and for silicon.
    }
    \label{fig:fig_2}
\end{figure}

\begin{figure}[t]
    \begin{center}
    \includegraphics[draft=false,keepaspectratio=true,clip, width=0.96\linewidth]{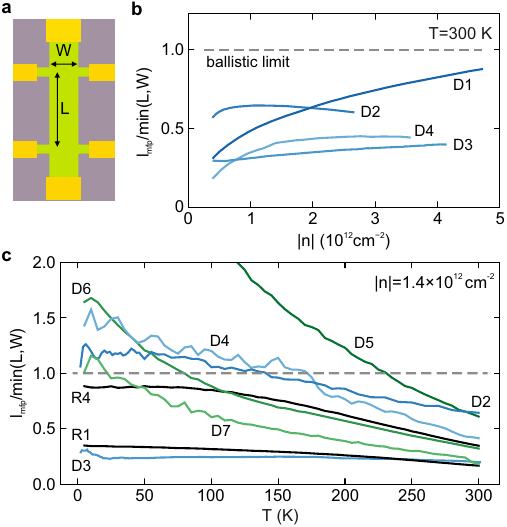}
    \end{center}
    \caption{\textbf{Mean free path and ballistic limit.}
    \textbf{a} Schematic illustration of the Hall bar device geometry with the distance between two voltage probes (length $L$) and the channel width $W$. % of the Hall bar.
    \textbf{b} The mean free path $l_{\mathrm{mfp}}$ divided by the smallest geometric extension of the system $min(L,W)$ as function of the charge carrier density ($n$) at $T= 300 \, \mathrm{K}$ for the thick WSe$_2$ devices (D1-D4).
    \textbf{c} The ratio $l_{\mathrm{mfp}}/min(L,W)$ as function of temperature at a charge carrier density of $|n| =  1.4 \times 10^{12} \, \mathrm{cm^{-2}}$. The grey dashed line is set to 1, the point at which the mean free path is equal to $min(L,W)$. %the minimum of $L$ and~$W$.
    }
    \label{fig:fig_3}
\end{figure}

\begin{figure*}[t]
    \begin{center}
    \includegraphics{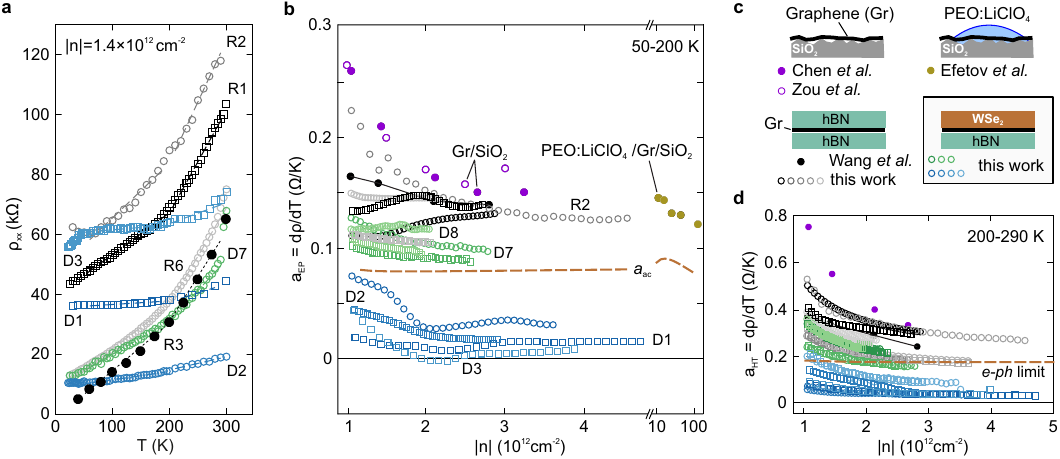}
    \end{center}
    \caption{\textbf{Suppression of the temperature-dependent resistivity of WSe$_2$/Gr/hBN heterostructures.} 
    \textbf{a} Resistivity as a function of temperature for four WSe$_2$/Gr/hBN devices (D1,D2,D3,D7) and four reference samples (R1-R3,R6). The dashed lines represent piece-wise linear fits in the temperature range of $50-200$~K and $200-290$~K. Open circles correspond to electron doping while open squares correspond to hole doping. R3 is taken from Wang et al.~\cite{Wang2013Nov}.
    \textbf{b} Slope ($a_\mathrm{EP}=d\rho/dT$) of the piece-wise linear fit highlighted in panel a in the range of 50-200~K for some devices (see labels). The brown dashed line corresponds to the theoretical expectation due to acoustic phonon scattering $a_\mathrm{ac}(\beta_\mathrm
    G)$ using the \textit{ab-initio} gauge-field coupling $\beta_\mathrm{G}$.
    Here also data from Chen et al.~\cite{Chen2008Mar}, Zou et al.~\cite{Zou2010Sep} and Efetov et al.~\cite{Efetov2010Dec} are included.
    \textbf{c} Schematics of the different sample types compared in panel a, b, and d. 
    \textbf{d} Slope ($a_\mathrm{HT}=d\rho/dT$) of the piece-wise linear fit from panel a in the range of 200-290~K for a set of devices (color corresponds to the labels in panels a and b).
    }
    \label{fig:fig_4}
\end{figure*}
\begin{figure*}[t]
    \begin{center}
        \includegraphics[draft=false,keepaspectratio=true,clip, width=1.0\linewidth]
        {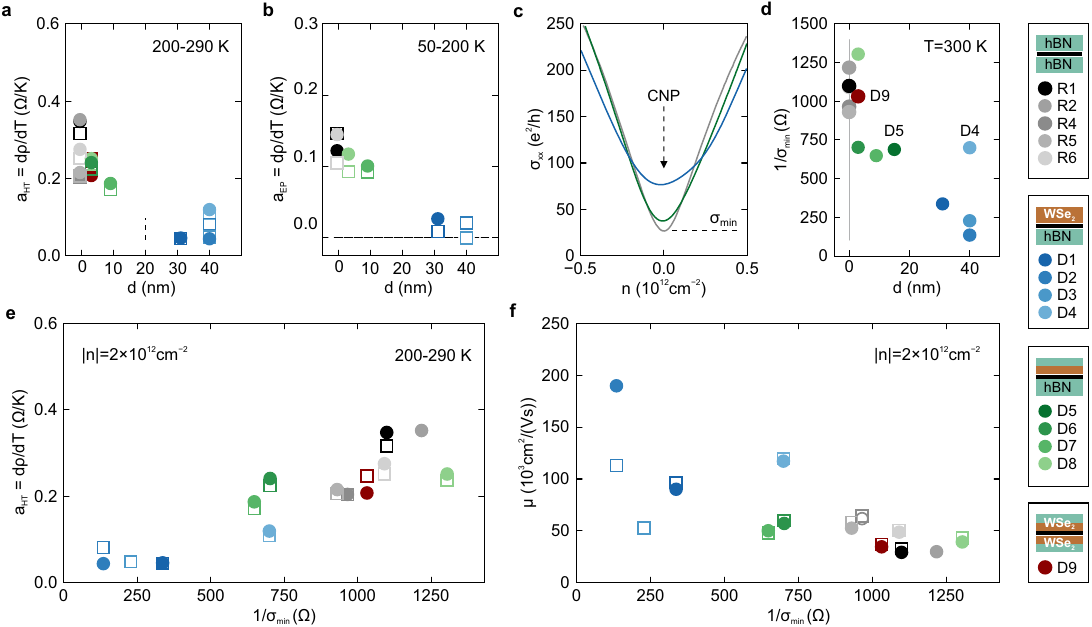}
    \end{center}
    \caption{\textbf{The role of the WSe$_2$ thickness and of charge disorder.}
    \textbf{a, b} Extracted slope of the temperature dependence $a_\mathrm{HT}$ in the range 200-290~K (panel a) and of $a_\mathrm{EP}$ in the range 50-200~K (panel b) as a function of the thickness $d$ of the WSe$_2$ layer for a fixed charge carrier density of $|n|=2\times10^{12}~$cm$^{-2}$. The reference devices hBN/Gr/hBN are set to a thickness of $d = 0$~nm. In all panels disks correspond to electron doping while open squares correspond to
hole doping. The labeling of the colored data points is summarized at the very right of the figure.
    \textbf{c} Conductivity as a function of charge carrier density close to the charge neutrality point (CNP) for three different devices at $T=$300~K, highlighting the value of the minimal conductivity $\sigma_{\mathrm{min}}$.
    \textbf{d} The $1/\sigma_\mathrm{{min}}$-value at room temperature as function of the thickness of the WSe$_2$ layer $d$. 
    \textbf{e} Slope of the temperature dependence $a_\mathrm{HT}$ in the range 200-290\,K as function of $1/\sigma_\mathrm{{min}}$ for $|n|=2\times10^{12}~$cm$^{-2}$.
    \textbf{f} Charge carrier mobility $\mu$ as a function of $1/\sigma_\mathrm{{min}}$ at room temperature and a charge carrier density of $|n|=\mathrm{2 \times 10^{12}\,cm^{-2}}$.
    }
    \label{fig:fig_5}
\end{figure*}

We estimate the charge carrier mobility $\mu$ at each charge carrier density using the linear relation $\sigma=e\mu n$. 
The room-temperature mobility of seven different WSe$_2$/Gr/hBN devices (D1-D7) is plotted in Fig.~\ref{fig:fig_2}, together with those of four reference samples based on hBN/Gr/hBN (R1-R4). 
Among these, R1 and R2 have the same Hall-bar geometry as D1-D3, while the data labelled as R3 corresponds to those originally published in Ref.~\cite{Wang2013Nov}, which is among the highest room temperature mobility value reported so far. 
At high carrier densities ($|n|>2\times 10^{12}$cm$^{-2}$), all reference devices show a room-temperature mobility of a few 10$^4$cm$^2$/(Vs), in below the theoretical predictions of phonon-limited conductivity in intrinsic graphene~\cite{Wang2013Nov,Park2014Mar,Sohier2014Sep}. 
This behavior has to be contrasted to that of the devices based on WSe$_2$/Gr/hBN. 
Devices D1, D2, and D4 show charge carrier mobilities that exceed the currently accepted upper limit for room-temperature mobility of graphene devices (see the grey dashed line in Fig.~\ref{fig:fig_2}). 
Devices D6 and D7, with WSe$_2$ thicknesses below 10~nm, remain below the electron-phonon (e-ph) limit. Only D5 (with around 15~nm thick WSe$_2$) slightly exceeds the limit at high charge carrier densities.
Device D3, despite having on a thick (40~nm) WSe$_2$ flake, shows reduced mobility, likely due to lower sample quality, as indicated by its low-temperature mobility of also just 50,000~cm$^2$/(Vs) (see SI).
These results suggest that achieving carrier mobilities beyond the current state of the art requires both sufficient WSe$_2$ thickness ($>$~15 nm) and high-quality devices.

To verify that the room temperature transport in our devices remains in the diffusive regime, we estimate the mean free path from the conductivity via $l_{\mathrm{mfp}} = h\sigma/(2e^2\sqrt{\pi n})$. 
We normalize this value by the smallest geometric dimension of the Hall bar, $min(L, W)$, where $L$ is the distance between voltage probes and $W$ the width of the Hall bar (see schematic in Fig.~\ref{fig:fig_3}a). 
A normalized value larger than 1 indicates that we are in the ballistic regime. 
As shown in Fig.~\ref{fig:fig_3}b, the resulting values confirm that all WSe$_2$/Gr/hBN devices based on thick WSe$_2$ remain in the diffusive regime at room temperature.
For the highest-mobility devices, $l_{\mathrm{mfp}}/min(L,W)$ approaches but does not exceed unity, validating the use of the Drude model for the mobility extraction even for the high-mobility devices.

To investigate the origin of the high room temperature mobilities observed in the WSe$_2$/Gr/hBN devices, we examined the temperature dependence of the resistivity. This provides insight into electron-phonon scattering~\cite{Sohier2014Sep} and electron scattering processes in general. 
Since extracting the electron-phonon coupling from the slope of the resistivity as function of temperature requires diffusive transport, we first track $l_{\mathrm{mfp}}/min(L,W)$ as a function of temperature at a fixed carrier density of $|n|=1.4\times 10^{12}$cm$^{-2} $ (see Fig.~\ref{fig:fig_3}c; traces of all devices are shown in the SI). 
For some devices (R1 and D3), the ratio $l_{\mathrm{mfp}}/\min(L,W)$ remains below unity over the full temperature range, confirming diffusive transport. 
Note that in devices with a high mobility and a small geometry, such as for example D4, the transport becomes ballistic below 175~K, as $l_{\mathrm{mfp}}/\min(L,W)$ exceeds unity. 
For the following analysis of the temperature-dependent resistivity, we therefore restrict the studied temperature range for each device to the regime where transport is diffusive. 

The resistivity as a function of temperature at a fixed charge carrier density ($|n|=1.4\times 10^{12}$cm$^{-2}$) is presented in Fig.~\ref{fig:fig_4}a for representative devices. 
Independent of the residual resistivity at low temperatures ($T \approx 2-5$~K), devices with thick WSe$_2$ layers (D1-D3; blue data points) show a much weaker temperature dependence compared to the reference devices and devices based on thinner WSe$_2$. 
To quantify this, we extract the slope of the resistivity as function of $T$ in two temperature intervals: (i) $a_\mathrm{EP} = d\rho/dT$ in the range $50-200$~K, the so-called \textit{equipartition} (EP) regime where in principle a linear increase in $\rho(T)$ is expected due to acoustic phonon scattering, and (ii) $a_\mathrm{HT} = d\rho/dT$, the \textit{high temperature} (HT) regime, in the range $200-290$~K, where optical phonon scattering is expected to become increasingly relevant~\cite{Sohier2014Sep}. 
The extracted slopes are shown as a function of charge carrier density in Figs.~\ref{fig:fig_4}b and~\ref{fig:fig_4}d, where we also compare with data from the literature for graphene on $\mathrm{SiO_2}$ with~\cite{Efetov2010Dec} and without electrolyte gating~\cite{Chen2008Mar,Zou2010Sep}. 
Reference samples show $a_\mathrm{EP}$ values of around 0.12~$\mathrm{\Omega/K}$ or larger, consistent with previous reports on hBN-encapsulated graphene~\cite{Wang2013Nov}. 
Devices with thin WSe$_2$ layers (green data points) partly approach a lower value of around 0.09~$\mathrm{\Omega/K}$. 
In stark contrast to both are WSe$_2$/Gr/hBN devices with thick WSe$_2$ (D1-D3; blue data points) which show values below $0.05 \, \mathrm{\Omega/K}$ and, in some cases even no temperature dependence. 
A similar trend is also observed in the high temperature regime, where $a_\mathrm{HT}$ is again strongly reduced for devices with thicker WSe$_2$. 
These results show that both acoustic and optical phonon scattering seem to be significantly suppressed in devices with thick WSe$_2$, pointing to a modification of the dominant scattering mechanism.

For the equipartition regime ($50-200$~K) at high carrier density ($|n|>2\times10^{12}\,$cm$^{-2}$), it is widely accepted that 
in clean samples the
resistivity is limited due to scattering with in-plane acoustic phonons mediated by gauge-field coupling, with only minor contributions from displacement-field coupling~\cite{Sohier2014Sep,Park2014Mar,Wallbank2018Oct}. 
Sample-dependent effects such as impurity scattering or umklapp electron-electron scattering~\cite{Wallbank2018Oct}, are expected to impact the total resistivity but not the temperature dependence. 
The expected linear temperature dependence can be approximated via
$\rho_{\mathrm{ac}} = 3\hbar \pi \beta_\mathrm{G}^2k_\mathrm{B} T /(4a_0e^2 \mu_{\mathrm{{Gr}}} v_\mathrm{A}^2) = a_{\text{ac}}\,T$,
 with the dimensionless gauge-field parameter $\beta_\mathrm{G}$, the effective acoustic sound velocity $v_\mathrm{A}$, graphene's mass density $\mu_\mathrm{{Gr}}$ and the lattice constant $a_0$ (for more details see SI)~\cite{Sohier2014Sep}. 
Using an \textit{ab initio} gauge-field parameter~\cite{Sohier2014Sep} yields a theoretically expected slope of approximately $a_\mathrm{ac} \approx 0.08 \, \mathrm{\Omega/K}$ (see orange dashed line in Figs.~\ref{fig:fig_4}b), in agreement with the values extracted for the reference devices and the WSe$_2$/Gr/hBN devices with a thin WSe$_2$. 

In contrast, explaining the significantly suppressed temperature dependence observed in devices incorporating thick WSe$_2$ layers (D1--D4) requires additional considerations. Such behavior would imply either (i) substantial modifications of the effective electron-phonon coupling strength in graphene, potentially arising from changes in the phonon spectrum or screening environment, or (ii) a change in the dominant momentum-relaxation mechanism, leading to a breakdown of a simple Matthiessen-type separation of scattering contributions.
At present, a detailed microscopic understanding of the origin of this behavior is still lacking. In particular, it remains unclear to what extent proximity-induced effects, disorder-assisted scattering processes, or substrate-mediated phonon interactions contribute to the observed suppression of the temperature dependence. 

We next turn to the high-temperature regime ($200-290$~K). 
Here, scattering with intrinsic optical zone-boundary phonons in graphene has been identified as the main intrinsic contribution, while a recent work points to a significant remote contribution from the ZO phonons of hBN in hBN-encapsulated devices~\cite{Dinar2026Apr}.
The electron-phonon coupling to these phonons is known to be sensitive to screening from charge carriers or the dielectric environment~\cite{Sohier2014Sep,Attaccalite2010Apr,Guandalini2025Apr,Moczko2025May}. 
The larger dielectric screening from the nearby WSe$_2$ layers can both reduce the coupling strength and lead to an increase in the involved phonon energies. 
These effects together could provide an explanation for the suppressed resistivity slope $a_\mathrm{HT}$ observed in Fig.~\ref{fig:fig_4}d. 
A comprehensive theoretical treatment is necessary to determine whether screening alone accounts for the observed suppression in the HT regime or if additional mechanisms are required.

The link between the observed effective suppression of phonon scattering in both discussed temperature regimes and the WSe$_2$ thickness might be related to a counter-intuitive role of charge disorder.
As shown in Figs.~\ref{fig:fig_5}a and~\ref{fig:fig_5}b, both slopes $a_\mathrm{EP}$ and $a_\mathrm{HT}$ decrease systematically with increasing WSe$_2$ layer thickness, suggesting that the suppressing mechanism is not limited to the WSe$_2$-graphene interface, but that the bulk WSe$_2$ plays an important role.
Figs.~\ref{fig:fig_5}c and d show that the minimal conductivity at the charge neutrality point (CNP) $\sigma_{\mathrm{min}}$ increases with WSe$_2$ thickness, corresponding to a reduced peak resistivity, $\rho_\mathrm{max}=1/\sigma_\mathrm{min}$ at the CNP. 
We note that measurements based on tuning a top gate (see Supplementary Information) confirm that this trend cannot be explained by the presence of a parallel conduction channel in WSe$_2$.
Instead, the observed increase in $\sigma_{\mathrm{min}}$ is consistent with enhanced charge disorder, which is expected to have its strongest impact in the vicinity of the charge neutrality point (CNP)~\cite{Chen2008May}. Notably, Fig.~\ref{fig:fig_5}e shows a systematic relation between the reduced temperature dependence, quantified by $a_\mathrm{HT}$ and $1/\sigma_{\mathrm{min}}$. This behavior suggests that increasing charge disorder possibly diminishes the role of temperature-dependent scattering mechanisms.
This trend also explains the apparent increase in room-temperature mobility shown in Fig.~\ref{fig:fig_5}f. A weaker temperature dependence means that the mobility is less strongly degraded at elevated temperature. Thus, within this regime, enhanced charge disorder possibly may counter-intuitively be associated with higher room-temperature mobility, not because disorder improves transport directly, but because it reduces the apparent impact of electron-phonon scattering.
Importantly, the room-temperature mobility is still constrained by scattering mechanisms that dominate at low temperatures. Consequently, even in devices such as D3, where phonon-related contributions appear strongly reduced, the room-temperature mobility does not exceed that of the reference devices, as it remains limited by residual disorder.

In summary, we observe that the room-temperature charge carrier mobility in graphene can be substantially enhanced in WSe$_2$/Gr/hBN heterostructures incorporating thick WSe$_2$ flakes. This enhancement is associated with a pronounced suppression of the temperature dependence of the resistivity in both the equipartition and high-temperature regimes. Notably, the reduced temperature dependence and the resulting increase in room-temperature mobility exhibit an unexpected relation to the minimum conductivity at the charge neutrality point, suggesting a connection to increased charge disorder and/or enhanced screening effects.

Our results are directly relevant for applications that require high-mobility graphene channels under ambient conditions, such as high-frequency electronics and Hall sensors. To further elucidate the microscopic origin of the observed behavior, future experiments could focus on identifying the relevant phonon modes. In the equipartition regime, wide-channel devices would enable the observation of magnetophonon oscillations, providing access to the sound velocities and energies of the participating acoustic phonon modes~\cite{Kumaravadivel2019Jul}. In the high-temperature regime, Raman spectroscopy using low excitation energies (1.16 eV or below) as a function of WSe$_2$  thickness could offer valuable insight into the role of screening in electron-phonon coupling and the phonon modes governing transport.

\section*{Data availability}
The data supporting the findings of this study are available in a Zenodo repository under 10.5281/zenodo.21156888
\section*{Acknowledgements}

The authors thank V. Falko, A. Geim and L. Wirtz for helpful discussions.
This work was supported by the European Union's Horizon 2020 Research and Innovation Program under grant agreement No. 785219, 881603 (Graphene Flagship) and from the European Research Council (ERC) under grant agreement No. 820254, the Deutsche Forschungsgemeinschaft (DFG, German Research Foundation) under Germany’s Excellence Strategy - Cluster of Excellence Matter and Light for Quantum Computing (ML4Q) EXC 2004/1 and  EXC 2004/2 - 390534769, by DFG through BE 2441/9-1, and by the Helmholtz-Nano-Facility (HNF)~\cite{Albrecht2017May}. K.W. and T.T. acknowledge support from the JSPS KAKENHI (Grant Numbers 21H05233 and 23H02052), the CREST (JPMJCR24A5), JST and World Premier International Research Center Initiative (WPI), MEXT, Japan. mputing resources were provided by PRACE on Marconi at CINECA, Italy (Project ID 2016163963).


% ============================================================
% arXiv include file generated from supplement.tex
% This file is NOT standalone; it is included by ms_arxiv.tex.
% The original supplement.tex is left unchanged.
% ============================================================

% Commands that existed only in the standalone SI preamble.
\newcommand{\bo}{\boldsymbol}
\newcommand{\boq}{\mathbf{q}}
\newcommand{\bok}{\mathbf{k}}
\newcommand{\bor}{\mathbf{r}}
\newcommand{\boG}{\mathbf{G}}
\newcommand{\boR}{\mathbf{R}}

\clearpage
\onecolumngrid
\linespread{1}\selectfont

% Reset counters so the merged SI behaves like the standalone SI.
\setcounter{section}{0}
\setcounter{subsection}{0}
\setcounter{subsubsection}{0}
\setcounter{figure}{0}
\setcounter{table}{0}
\setcounter{equation}{0}
\setcounter{footnote}{0}

\renewcommand{\thesection}{S\arabic{section}}
\renewcommand{\thefigure}{S\arabic{figure}}
\renewcommand{\thetable}{S\arabic{table}}
\renewcommand{\theequation}{S\arabic{equation}}

% Unique hyperref destinations after resetting the SI counters.
\providecommand{\theHsection}{}
\providecommand{\theHsubsection}{}
\providecommand{\theHsubsubsection}{}
\providecommand{\theHfigure}{}
\providecommand{\theHtable}{}
\providecommand{\theHequation}{}
\renewcommand{\theHsection}{SI.section.\arabic{section}}
\renewcommand{\theHsubsection}{SI.subsection.\arabic{section}.\arabic{subsection}}
\renewcommand{\theHsubsubsection}{SI.subsubsection.\arabic{section}.\arabic{subsection}.\arabic{subsubsection}}
\renewcommand{\theHfigure}{SI.figure.\arabic{figure}}
\renewcommand{\theHtable}{SI.table.\arabic{table}}
\renewcommand{\theHequation}{SI.equation.\arabic{equation}}

% Preserve the standalone SI reference presentation [S1], [S2], ...
\renewcommand{\citenumfont}[1]{S#1}
\renewcommand{\bibnumfmt}[1]{[S#1]}

% REVTeX cannot produce a second independent front-matter block after the
% manuscript \maketitle. The full author list is already on page 1, so the
% appended SI starts with its title here.
\begin{center}
{\Large\bfseries Supplementary Information}\\[0.8em]
{\large Extraordinary high room-temperature carrier mobility in graphene-WSe$_2$ heterostructures}
\end{center}
\vspace{1em}

\section{Supplementary Methods}

\subsection{Characterization of the WSe$_2$/Gr/hBN devices}

\subsubsection{Scanning transmission electron microscopy measurements}

% STEM setup/measurement
A cross-sectional sample of one of the devices (device D1) was prepared by focused ion beam (FIB). High angle annular dark-field (HAADF) images were acquired in scanning transmission electron microscopy (STEM) at (a) a FEI Titan 80-300 STEM \cite{titan_s_2016} operated at 200~keV and (b) a FEI Titan G2 80-200 \cite{chemistem_2016} operated at 80~kV. Both microscopes are equipped with hardware aberration correctors for the probe-forming lens. The sample was first investigated at 80~kV (settings (a)) in order to avoid irradiation damage during the acquisition of energy-dispersive X-ray spectra employing a chemiSTEM system. Further HAADF images have been acquired at 200~kV (settings (b)), providing a slightly better spatial resolution for the structural analysis. The HAADF image in Fig.~1b in the main text was acquired with settings (b) and a frame time of 17~s, collecting all angles from 35 to 230~mrad. Because of the high contrast differences arising from the presence of very light (hBN, graphene) and rather heavy elements (WSe$_2$), a band pass filter and a Gaussian filter were applied. This yields the local contrast variation due to the atomic lattice planes relative to a common background level. Additionally, separate color maps are used above and below the interface between graphene and WSe$_2$ to further enhance the lattice plane contrast.

The corresponding line profile was obtained after averaging the original image along the vertical axis, followed by band pass filtering. The original image is shown on a gray scale in Fig.~\ref{fig:s12}. The peak positions in the line profile were determined from Gaussian fits. This procedure was repeated for four individual images acquired with both experimental settings. From these four measurements, we determined the mean spacing between the graphene layer and the closest WSe$_2$ layer to be $5.3\pm0.3$~\AA. The uncertainty corresponds to the standard deviation of these measurements. An overview of the full layer sequence is displayed in Fig.~\ref{fig:s13}A in form of an HAADF image that was averaged along the  vertical direction. Additionally, electron diffraction x-ray (EDX) spectrum images were acquired using settings (b) simultaneously with the HAADF signal shown in Fig.~\ref{fig:s13}A, for a total frame time of 33~minutes using sequential frames and a drift correction. 

% EDX
The EDX spectrum image was averaged along the vertical direction, yielding a two dimensional data set, i.e., one spectrum for every horizontal scan position. Principal component analysis (PCA) of the EDX data was performed using the non-negative matrix factorization, including seven components, as implemented in the HyperSpy software package~\cite{hyperspy_2017}. The resulting distribution of the individual components score is displayed in Fig.~\ref{fig:s13}B, which can be related directly to the HAADF image above. Component no. 6 displays the carbon signal, which is peaked sharply  at the position between hBN and WSe$_2$, indicating the presence of a graphene layer. The enhancement of C at the other interfaces (SiO$_2$/hBN and WSe$_2$/hBN) is less pronounced and most probably related to contamination of these interfaces. The Cu signal originates from the TEM copper grid that holds the specimen. The spectra of all seven components are displayed separately in Fig.~\ref{fig:s14}.

\subsubsection{Tunability of the charge-carriers density in graphene in WSe$_2$/Gr/hBN devices}\label{sec:tunability}

We performed a series of different experiments to investigate the contribution of the WSe$_2$ layer to the conductivity of the WSe$_2$/Gr/hBN devices and the tunability of the charge-carriers density in graphene. In first place, we have fabricated a device with a top and a back gate (device D1, Fig.~1c,d in the main text). The two gates allow tuning of the Fermi energy in WSe$_2$ independently of the one in graphene. The room-temperature (300~K) conductivity of D1 is shown in Fig.~\ref{fS1}A as function of  top- and back-gate voltage. The top gate tunes the device in the range V$ _\mathrm{TG}$~=~-2.5~ to 0.5~V, where the Fermi energy of WSe$_2$ lies in the band gap. For larger values of  V$ _\mathrm{TG}$, electrons start occupying the  conduction band of WSe$_2$, screening graphene from the top gate and suppressing the top-gate tunability of the device. Notably, the resistivity of the device does not change significantly when this occurs (see Fig.~\ref{fS1}B), indicating that the WSe$_2$ layer represents only a weak parallel conduction channel that is not gate tunable. This observation is consistent with previous reports on high contact resistances and low charge carrier mobilities in WSe$_2$ \cite{Fang2012Jul}.

\subsubsection{Quantum Hall measurements}

To further confirm that the transport properties of the devices are dominated by graphene we performed quantum Hall-effect measurements at 100~mK using magnetic fields of up to 14~T, see e.g. Fig.~\ref{fS2}A-B for devices D1-D3. The distinct plateaus of the transverse conductivity, $\sigma_{\mathrm{xy}}$ at  filling factors $\nu=-10,-6,-2,2$ and 6 (see Fig.~\ref{fS2}A), as well as the Shubnikov-de-Haas oscillations of $R_\mathrm{xx}$ (inset of Fig.~\ref{fS2}A), are a clear fingerprint that graphene dominates the transport properties of the device. A small contribution from WSe$_2$ becomes visible at high magnetic fields as an overshoot of $\sigma_{\mathrm{xy}}$ close to the charge neutrality point ($n=0$~cm$^{-2}$). Quantum Hall-effect measurements on devices D2 and D3 similarly indicate that the transport properties of these devices are dominated by graphene (see Figs.~\ref{fS2}C and~\ref{fS2}D). 

In Fig.~\ref{QH_2}, we additionally show the quantum Hall measurements of all other devices (the reference hBN/Gr/hBN devices R4-R6 and the WSe$_2$/Gr/hBN devices D4-D9) discussed in this work. These additional measurements were performed at 2~K with magnetic fields up to 7~T. For all devices we observe the formation of Landau levels similar to devices D1-D3 shown in Fig.~\ref{fS2}. The degree of visibility of the minima in the resistivity $\rho_\mathrm{xx}$ varies between the different devices, but for all devices, we can clearly follow the position of the minima in the maps, which we need for the extraction of the lever arm from these measurements as discussed in the next paragraph.

To further support the observation that the WSe$_2$ is not contributing to the conductivity, we extracted the back-gate lever arm, $\alpha_\mathrm{HE}$, by standard Hall-effect measurements at 2~K, see Figs.~\ref{hall}A and~\ref{hall}B. We compare this quantity with the lever-arm, $\alpha_{\rm QHE}$, determined from quantum Hall-effect measurements by fitting the relation $B_{\nu}(V_{\rm BG})=\alpha_{\rm QHE} h (V_{\rm BG}-V_{\rm BG}^0)/2e(2\nu+1)$ to the position of the minima of $R_{xx}$ in the Landau fan measurements (dashed lines in Figs.~\ref{fS2}B,~\ref{fS2}C and~\ref{fS2}D).  Here, $\nu$ is the filling factor. The lever-arm extracted from the half-integer quantum Hall effect $\alpha_{\rm QHE}$ depends only on the charge carriers in graphene, while $\alpha_\mathrm{HE}$ is sensitive to charge carriers both in graphene and in WSe$_2$. For all devices, for which $\alpha_\mathrm{HE}$ was extracted, we observe a very good agreement between these two values (see Table~\ref{tab:alpha}),  indicating that the small parallel conduction-channel provided by WSe$_2$ is not tuned by $V_{\rm BG}$ and, as such, it does not affect the dependence of the conductivity $\sigma$  on the carrier density of graphene $n$. Furthermore, $\alpha_{\rm QHE}$ and $\alpha_{\rm HE}$ agree also well with the estimate of the lever arm based on a plate capacitor model and the geometry of the sample $\alpha_{\rm geom}=7.4\times10^{10}$~cm$^{-2}$~V$^{-1}$.  

%% lever arm values used
We consistently use the lever-arm extracted from quantum Hall-effect measurements to convert back-gate voltages to charge-carrier densities in graphene, according to the relation 
\begin{equation}
n=\alpha_{\rm QHE} (V_{\rm BG}-V_{\rm BG}^0) = \alpha_{\rm QHE} V_{\rm BG}+n_0,
\end{equation}
with the values of $\alpha_{\rm QHE}$, $V_{\rm BG}^0$ and $n_0$ given in Table~\ref{tab:alpha}. In principle, the position of the charge neutrality point $V_{\rm BG}^0$ could depend on temperature if there is temperature-activated charge transfer between WSe$_2$ and graphene. However, we rule out this possibility by temperature-dependent measurements that show that the charge neutrality point (CNP) does not appreciably shift over the whole temperature range, see Fig.~\ref{hall}C,D.

\subsubsection{Conductivity of all measured devices at room temperature}
The conductivity of all measured devices (D1--D8, R1,R2,R4--R6) at room temperature (300~K) is plotted in Figs.~\ref{fS3}A and~\ref{fS3}B as a function of charge carrier density. The conductivities of the five different reference devices vary, which is likely due to different amount of extrinsic disorder, such as strain variations, but never exceed a value of $\mathrm{600\,e^2/h}$, and therefore stay below the phonon-limit (see main text). The devices based on WSe$_2$/Gr/hBN show a wider spread of conductivities (blue traces) and exceed to much larger values for the devices with thick WSe$_2$ (D1-D4). The asymmetry between the electron and  hole transport results from doping inhomogeneities in the samples.   
Figure~\ref{fS3}C shows the conductivity of device D3, both at low (2~K, green trace) and at room temperature (300~K, blue trace). 
The mobility of this sample increases little at low temperature, indicating that its transport properties are strongly dominated by impurity scattering.

\subsubsection{Mean free path in WSe$_2$/Gr/hBN devices: ruling out ballistic transport at elevated temperatures}

In Fig.~\ref{mean-free-path} the mean free path normalized to the smallest geometric length of all samples, of which the temperature dependence is discussed are presented. For the discussion of the temperature dependence of the resistivity only the samples and temperature regions in which the mean free path is below unity are discussed. From Fig.~\ref{mean-free-path}C it becomes evident that this condition is true for all thick WSe$_2$ devices, which show an increased charge carrier mobility in comparison to hBN-encapsulated devices.

\subsubsection{Long term stability of the WSe$_2$/Gr/hBN devices}
Figure~\ref{long-time-stability} shows two measurements of the room-temperature charge carrier mobility of device D1, taken one year apart (green and blue traces). The excellent agreement between the two data-sets indicates high robustness and good long term stability of the device, which is a necessary requirement for future technological applications.

\section{Supplementary Discussion}

\subsection{On the relation between $\mu$ and $n^*$}\label{sec: discussion-n*}
\label{sub:mu_n*}
A striking difference between standard high-mobility hBN/Gr/hBN devices and devices based on WSe$_2$/Gr/hBN is that the latter present high mobilities {\em and} relatively broad and low resistance peaks around charge neutrality point (see e.g. Fig.~\ref{hall}D), while standard hBN/Gr/hBN devices show an approximately linear relation between the inverse mobility, $1/\mu$, and width of the resistance peak, $n^*$. 

This relation has been investigated extensively by Couto {\em et al.}~\cite{Couto2014Oct} for graphene on hBN devices. This work provides strong evidence that random local strain variations in the graphene lattice are the main cause of scattering for charge carriers in high-quality graphene devices. %, and therefore the ultimate limitation to the mobility. 
Random local strain fluctuations in the graphene lattice can be probed by Raman spectroscopy, as the width of the Raman 2D peak is highly sensitive to strain inhomogeneities on length scales smaller than the laser spot ($<500$~nm)~\cite{Neumann2015Sep}. High-mobility graphene samples are characterized by narrow Raman 2D peaks with full-width-at-half-maximum around 20 cm$^{-1}$ or below (as  in the case of the WSe$_2$/Gr/hBN devices).
%, see Sec.~\ref{sec:raman1}).

The effect of random strain variations on the motion of electrons in graphene can be described by introducing a scalar and a vector potential in the long-wavelength Dirac Hamiltonian~\cite{Vozmediano2010Nov}.  The scalar potential induces random local variations of the charge density and affects the  residual charge density $n^*$.  The vector potential can be thought of as a random pseudo-magnetic field and represents the main source of electron scattering in graphene~\cite{Couto2014Oct}.\footnote{We note in passing that this conclusion is fully consistent with what emerges from first-principles calculations of the electron-phonon scattering in graphene~\cite{SI:Sohier2014Sep}, which indicate that electrons couple to the phonons dominantly through the so-called gauge-field, with only marginal contributions from the scalar deformation potential.}
The linear relation between $1/\mu$ and $n^*$ observed in  graphene on hBN (and also on SiO$_2$ and on SrTiO$_3$) stems from the fact that in these systems both quantities are related to the correlation function of the random strain field~\cite{Couto2014Oct}, the former through vector-potential, the latter through scalar-potential.

The situation can however be different for graphene on other substrates, since  mechanisms that contribute to the scalar but not to the vector potential (e.g. charge disorder) affect $n^*$  but have little impact on $\mu$  at large charge-carrier density.  
WSe$_2$ is known to be affected by considerable charge disorder~\cite{Rhodes2019May}. The potential fluctuations caused by this charge disorder add to the scalar potential caused by the random strain-field and result in a much broader resistance peak for the WSe$_2$/Gr/hBN devices than what one would expect for graphene devices of comparable mobility on hBN.
This is the reason why the WSe$_2$/Gr/hBN devices don't show the linear dependence between $1/\mu$ and $n^*$ observed in standard high-mobility devices.

Charge disorder in the WSe$_2$ layer is efficiently screened by the graphene's charge carriers at high carrier density, but it affects the behavior of the devices in all the situations where the density of states of graphene goes to zero, such as around charge neutrality point or in the presence of a Landau gap. This is why the WSe$_2$/Gr/hBN devices show no degeneracy lifting of the Landau levels (see Fig.~\ref{fS2}) despite their high mobility.
Charge disorder in WSe$_2$ can also be screened by charged carrier in the WSe$_2$ itself. Figure S\ref{fig:screening} shows measurements on device D1 similar to those presented in Fig.~\ref{fS1} but performed at 2~K. These clearly show how the resistivity peak of the device gets higher and sharper when the valence band of WSe$_2$ is populated and holes can (at least partially) screen the charge disorder present in this material. The same effect has been observed also in graphene on MoS$_2$ devices~\cite{Banszerus2017Jul}.

\subsection{Approximate expression for the resistivity in the equipartition regime}

In the equipartition regime ($50-200$~K), optical phonons  contribute little to the electron-phonon scattering and it is possible to derive an approximate analytical expression for the contribution of the acoustic phonons to the resisitivity of graphene. The derivation below follows closely the one given in appendix C of Ref.~\cite{SI:Sohier2014Sep}. It is presented here to show explicitly how the increase of the resisitivty as a function of temperature depends on the phonon dispersion relation.

In the equipartition regime ($k_BT\ll \varepsilon_F$), the scattering with  longitudinal (LA) and transverse (TA) acoustic phonons can be considered elastic, and its contribution  to the resisitivity of graphene can be written as~\cite{SI:Sohier2014Sep}
\begin{align}
    \rho=\frac{2}{e^2v_F^2{\rm DOS}(\varepsilon_F)\tau_{\rm A}(\varepsilon_F)},
\end{align}
where ${\rm DOS}(\varepsilon_F)=2\varepsilon_F/\pi(\hbar v_F)^2$ is the density of states at the Fermi energy $\varepsilon_F$, and $\tau_{\rm A}$ is the scattering time due to acoustic phonons:
\begin{align}\label{eq:tauA}
\frac{1}{\tau_{\rm{A}}(\varepsilon_{\mathbf{k}})}  &=
\sum_{\mathbf{k'}}  P_{\mathbf{kk'},\rm A}
 (1-\cos(\theta_{\mathbf{k'}}-\theta_{\mathbf{k}})).
\end{align}
Here
\begin{align}\label{eq:Pkk}
P_{\mathbf{kk'},\rm A} &\approx
 \frac{2\pi}{\hbar} \frac{1}{N}  \sum_{\nu=\rm TA,LA}
|g_{\mathbf{k'},\mathbf{k},\nu}|^2  \left[2
n_{\nu,\mathbf{k'}-\mathbf{k}}+1]
\delta(\varepsilon_{\mathbf{k'}}-\varepsilon_{\mathbf{k}}
\right)
\end{align}
is the scattering probability due to acoustic phonons between electronic states with momentum $\bok$ and $\bok'$, $\theta_{\bok(\bok')}$ and $\varepsilon_{\bok(\bok')}$ are the angle and the energy associated with the electronic momentum $\bok\, (\bok')$,  $N$ is the number of electronic states in the Brillouin zone,
$n_{\nu,\bok'-\bok}$ is the population of phonon mode $\nu$ at momentum $\bok'-\bok$ at equilibrium (Bose-Einstein distribution), %at energy $\hbar \omega_{\boq,\nu}$), 
and $\left|g_{\mathbf{k'},\mathbf{k},\nu}\right|^2$ is the modulus squared of the electron-phonon matrix element. In the limit of small phonon-momentum ($|\bok'-\bok|\to0$), the electron-phonon matrix element can be written as \cite{Manes2007Jul,SI:Sohier2014Sep}
\begin{align}\label{eq:matrixel}
    \left| g_{\mathbf{k'},\mathbf{k},\nu}\right|^2=\frac{\hbar}{2M\omega_{\nu,\bok'-\bok}}\frac{3 (\hbar v_F)^2}{8a^2_0} \beta_{\rm{G}}^2|\bok'-\bok|^2\frac{1\pm\cos(4\theta_{\bok'-\bok}+\theta_{\bok'}+\theta_{\bok})}{2},
\end{align}
where the $\pm$ sign stands for $\nu = $ LA and TA modes, respectively. Here, $M$ is the mass of the carbon atom,  $a_0=2.46$ \AA\ is graphene's lattice parameter,  $\omega_{\nu,\bok'-\bok}$ is the phonon frequency,  and $\beta_{\rm G}$ is the dimensionless acoustic gauge field, defined as in Ref.~\cite{Bistoni2019Jul}:
\begin{equation}\label{eq:betaG}
    \beta_{\rm{G}} = \frac{2\sqrt{2}a_0}{\sqrt{3}\hbar v_F} \beta_A,
\end{equation}
 where $\beta_A$ is the acoustic gauge field strength defined in Refs.~\cite{SI:Sohier2014Sep},\footnote{The gauge-field contribution to electron-phonon coupling is due to the displacement of the Dirac cone in k-space caused by the displacement of the atoms according to the phonon pattern. The displacement of the cone is independent of the Fermi velocity and robust to corrections from electron-electron interactions~\cite{SI:Sohier2014Sep}.
The parameter $\beta_G$ is directly proportional to this displacement. The parameter $\beta_A$ initially used in Ref~\cite{SI:Sohier2014Sep}, represents instead the opening of a gap if one stays at a fixed electronic momentum while the cone displaces. As such, it is proportional to the Fermi velocity. We chose to use $\beta_G$ in this work because it is more fundamental and highlights the independence of the resistivity on the Fermi velocity.}
In Eq.~\eqref{eq:matrixel}, we neglected the deformation potential, whose effect was found to be negligible with respect to the acoustic gauge-field %term $\beta_{\rm{G}}$
\cite{SI:Sohier2014Sep}.
The expression for the scattering probability, Eq.~\eqref{eq:Pkk},  can be further simplified as follows:
\begin{align}\label{eq:Pkk-1}
P_{\mathbf{kk'},\rm A} &\approx
 \frac{\pi}{N} \frac{3 (\hbar v_F)^2}{8a^2_0} \frac{\beta_{\rm G}^2}{M}\frac{k_BT}{\hbar}  \sum_{\nu=\rm TA,LA}\frac{
|\mathbf{k'}-\mathbf{k}|^2}{\omega^2_{\nu,\bok'-\bok}}\left[1\pm\cos(4\theta_{\bok'-\bok}+\theta_{\bok'}+\theta_{\bok})\right]\delta(\varepsilon_{\bok'}-\varepsilon_{\bok}) \nonumber\\
&\approx\frac{\pi}{N} \frac{3 (\hbar v_F)^2}{8a^2_0} \frac{\beta_{\rm G}^2}{M}\frac{k_BT}{\hbar} \frac{
2|\mathbf{k'}-\mathbf{k}|^2}{\omega^2_{A,\bok'-\bok}}\delta(\varepsilon_{\bok'}-\varepsilon_{\bok}),
\end{align}
where we have taken into account that  $n_{\nu,\bok'-\bok} \approx\frac{k_BT}{\hbar\omega_{\nu,\bok'-\bok}}\gg1$ in the considered temperature regime and, in the second line, we also made the approximation $\omega_{\rm LA,\boq}\approx \omega_{\rm TA,\boq}\approx \omega_{\rm A,\boq}$.
Taking into account Eq.~\eqref{eq:tauA} and Eq.~\eqref{eq:Pkk-1}, the resisitivity is then given by
\begin{equation}\rho=a\left(\beta_{\rm G},\omega_{\rm A, \boq} \right)\cdot T, 
\end{equation}
with
\begin{align}
    a\left(\beta_{\rm G},\omega_{\rm A, \boq} \right)=&\frac{\pi^2\hbar}{e^2\varepsilon_F} \frac{3 (\hbar v_F)^2}{8a^2_0}
    \frac{\beta_{\rm{G}}^2k_B}{M}\frac{1}{N}\sum_{\bok'}\frac{2|\bok'-\bok|^2}{\omega^2_{{\rm A}, \bok'-\bok}}\delta(\varepsilon_{\bok'}-\varepsilon_{\bok})(1-\cos(\theta_{\bok'}-\theta_{\bok})).
\end{align}
Replacing the sum over momenta with an integral over energy,
\begin{align}
\frac{1}{N} \sum_{\bok'}= \frac{S_{\Re}}{(\hbar v_F)^2}
\int \frac{\varepsilon_{\bok'}d\varepsilon_{\bok'} d\theta_{\bok'}}{(2\pi)^2},
\end{align}
and taking into account that $\bok$ and $\bok'$ are on an isoenergetic line with energy $\varepsilon_F$, we  arrive at the final expression for the slope of the resistivity as a function of temperature:
\begin{align}\label{eq:rho-approx}
    a\left(\beta_{\rm G},\omega_{\rm A, \boq} \right)=& \frac{3 \hbar k_B \beta_{\rm{G}}^2}{2a_0 e^2 \mu_{\rm Gr}}\cdot \int_{-\pi/2}^{\pi/2}d\theta_{\boq}\frac{\left(2k_F\cos^2\theta_{\boq}\right)^2}{\omega_{{\rm A}, 2k_F |\cos\theta_{\boq}|}^2},
\end{align}
with $\boq=\bok'-\bok$ and  $\mu_{\rm Gr}=2M/S_\Re$ the mass surface-density of graphene, being $S_\Re$  the unit-cell area.  

For the case of the linear dispersion relation  $\omega_{\rm A, \boq}=v_{\rm A}|\boq|$ this expression gives
\begin{align}\label{eq:rho-approx-1}
    \rho=& \frac{3\hbar\pi\beta_{\rm G}^2k_B}{4a_0 e^2 \mu_{\rm Gr}v_{\rm A}^2}\cdot T,
\end{align}
which is the approximate solution derived in Ref.~\cite{SI:Sohier2014Sep} for the case of isolated graphene, with $v_{\rm A}$ the effective sound velocity of the acoustic branch , $v_{\rm A}^{-2}=(v_{\rm TA}^{-2}+v_{\rm LA}^{-2})/2$. In the presence of shear modes, the quadratic dispersion relation  $\omega_{\rm A,\boq}=\sqrt{\omega_\Gamma^2+v_{\rm A}^2 |\boq|^2}$ can lead to a substantial suppression of $a$.

%%%%%%%%%%%%%%%%%%%%%

\section{Supplementary Tables}
\begin{table}[!htb]
\caption{\small Numerical values of the back-gate lever arm extracted  via quantum Hall-effect measurements, $\alpha_{\rm QHE}$, and via standard Hall- effect measurements at a temperature of 2~K, $\alpha_{\rm HE}$. Also shown are the position of the charge neutrality point, $V^0_{\rm BG}$, and the overall doping in graphene ${n_0=-\alpha_{\rm QHE} \, V_{\rm BG}^0}$ }
\label{tab:alpha}
\begin{center}
\begin{tabular}{ |c|c|c|c|c|}
\hline
Sample & $\alpha_{\mathtt{QHE}}$ ($10^{10}$~cm$^{-2}$V$^{-1}$) &  $\alpha_{\mathtt{HE}}$ ($10^{10}$~cm$^{-2}$V$^{-1}$) &  $V_{\mathtt{BG}}^0$ ($V$) &  $n_{\mathtt{0}} (10^{11}$~cm$^{-2}$) \\
\hline
D1 & 6.7 &  6.9& 5.3 & -3.6 \\
\hline
D2 & 6.9 & 7.1 & 8.2 & -5.6 \\
\hline
D3 & 6.7 & 7.1  & 1.1& -0.7 \\
\hline
D4 & 7.1 & - & -4.49 & 3.18 \\
\hline
D5 & 7.0 & 7.0 & -0.87 & 0.61 \\
\hline
D6 & 5.6 & 5.8 & 1.01& -0.57 \\
\hline
D7 & 6.6 & 6.9 & -1.81 & 1.20 \\
\hline
D8 & 6.5 & 6.6 &-0.89 & 0.58 \\
\hline
D9 & 6.6 & 6.6 & -1.59 & 1.05 \\
\hline
\hline
R1 &6.9 & 7.0 & 3.2 &  -2.2 \\
\hline
R2 & 7.0 & 7.1 & 3.2& -2.2 \\
\hline
R4 & 6.5 & -  &-0.7 & 0.45\\
\hline
R5 & 6.1 & 6.1 & -0.01 &0.01 \\
\hline
R6 & 6.9 & - & -0.91 & 0.63 \\
\hline

\end{tabular}
\end{center}
\end{table}

\begin{table}[!htb]
\caption{\small Thickness of the WSe$_2$ ($d$), Width ($W$), distance between voltage probes ($L$) and total length ($L_\mathrm{tot}$) of the Hall-bar devices.}
\label{tab:geometry}
\begin{center}
\begin{tabular}{ |c||c|c|c|c|}
\hline
 Sample & $d$~(nm) & $L$~($\mu$m) &  $W$~($\mu$m) & hBN source \\
\hline
\hline
D1 & 31 & 3.5 & 2.5 & HPHT-NIMS \\
\hline
D2 & 40 & 6.5 & 5 & HPHT-NIMS \\
\hline
D3 & 40 & 3.5 &  2.5 & HPHT-NIMS \\
\hline
D4 & 40 & 4.4 & 6 & APHT-RWTH \\
\hline
D5 & 15 & 2 & 2 & APHT-RWTH \\
\hline
D6 & 3 & 3 & 3 & HPHT-NIMS \\
\hline
D7 & 9 & 5 & 13 & APHT-RWTH \\
\hline
D8 & 3 & 1.6 & 3 & APHT-RWTH \\
\hline
D9 & 3/3 & 2 & 3.7 & APHT-RWTH \\
\hline
\hline
R1 & - & 5 & 3  & HPHT-NIMS \\
\hline
R2 & - & 3 & 2 & HPHT-NIMS \\
\hline
R4 & - & 3 & 2 & APHT-RWTH \\
\hline
R5 & - & 3 & 2 & APHT-RWTH\\
\hline
R6 & - & 3 & 2 & HPHT-NIMS \\
\hline

\end{tabular}
\end{center}
\end{table}

\begin{table}[!htb]
\caption{\small Numerical values of the parameters used for the calculations.
Here, u is the atomic mass unit. The dimensionless acoustic gauge-field parameter is defined as $\beta_{\rm{G}} = 2\sqrt{2}a_0 \beta_A/(\sqrt{3}\hbar v_F)$, where $\beta_A$ is the parameter calculated by first principles in Ref.~\cite{SI:Sohier2014Sep}.}
\label{tab:Sim_param}
\begin{center}
\begin{tabular}{ l c c c }
\hline
Parameter & Symbol  &Value  \\
\hline
Acoustic gauge field parameter & $\beta_{\rm{G}}$ & $2.64$ \\
Lattice parameter & $a_0$ & $2.46$ \AA \\
Unit-cell area & $S_\Re$ & $5.24$ \AA$^2$ \\
Sound velocity TA & $v_{\rm{TA}} $ & $13.6$ km/s \\
Sound velocity LA & $v_{\rm{LA}} $ & $21.4$ km/s \\
Effective sound velocity & $v_{\rm A}=(v_{\rm TA}^{-2}+v_{\rm LA}^{-2})^2/4$ & $16.23$ km/s \\
Carbon atom mass &  $M$ & $12.0107$ u   \\
Mass density     & $\mu_\text{Gr}=2M/S_\Re$ &  $7.66$ kg/m$^2$    \\
Fermi velocity (GW) & $v_F$ & $1.00$ $10^6$ m/s \\
 \hline
\end{tabular}
\end{center}
\end{table}

\clearpage
%\newpage
\makeatletter
\let\SI@origlabel\label
\let\SI@origref\ref
\def\SI@lastbib{LastBibItem}
\renewcommand{\label}[1]{%
  \def\SI@tmp{#1}%
  \ifx\SI@tmp\SI@lastbib
    \SI@origlabel{LastBibItemSI}%
  \else
    \SI@origlabel{#1}%
  \fi
}
\renewcommand{\ref}[1]{%
  \def\SI@tmp{#1}%
  \ifx\SI@tmp\SI@lastbib
    \SI@origref{LastBibItemSI}%
  \else
    \SI@origref{#1}%
  \fi
}
\makeatother

\newpage

\begin{figure}[t!]
    \centering
    \includegraphics[width=0.4\textwidth]{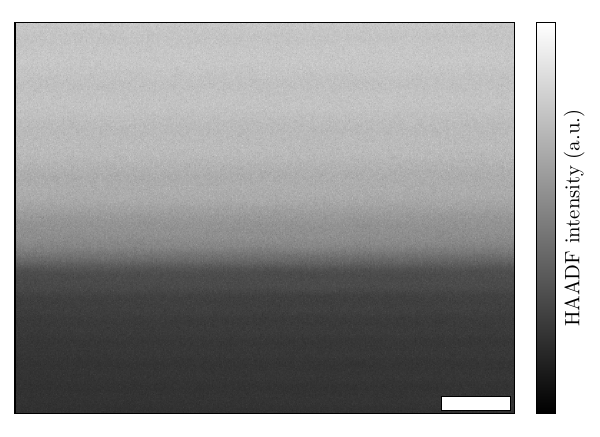}
    \caption{\small High-resolution high-angle annular dark-field (HAADF) image of the hBN/graphene/WSe$_2$ interface. The intensity increases rapidly at the graphene/WSe$_2$ interface, as the HAADF signal is proportional to the atomic number of the material. For this reason, the color scales in Fig.~1c of the main text have been adjusted for best visibility of the lattice fringes. The scale bar is 1~nm. }
    \label{fig:s12}
    \end{figure}
    
\begin{figure}[h!]
    \centering
    \includegraphics[width=0.8\textwidth]{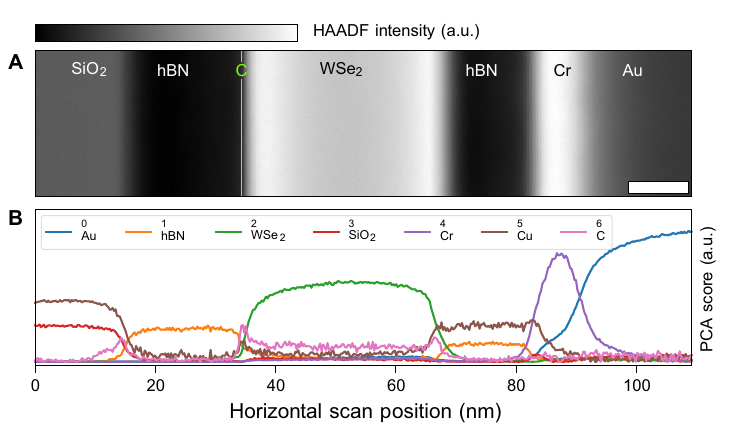}
    \caption{\small 
    {\bf A} HAADF image of the full layer sequence prepared by focused ion beam. The different materials are labeled and a green line marks the approximate position of the graphene layer. The scale bar is 10~nm. 
    {\bf B} Score of  principal component analysis (PCA). Each component is plotted as a function of horizontal scan position and is labeled by the most dominant material within the respective component.}
    \label{fig:s13}
    \end{figure}

    \begin{figure}[h!]
    \centering
    \includegraphics[width=0.8\textwidth]{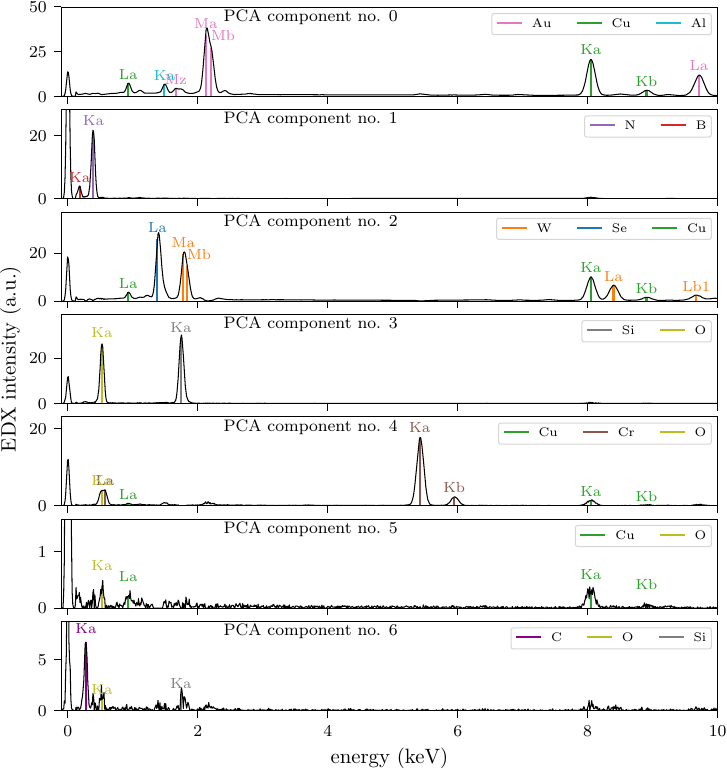}
    \caption{\small EDX spectra of all seven components determined by PCA. The most dominant x-ray lines in every spectrum are drawn as colored vertical lines (as labeled in the legend).}
    \label{fig:s14}
    \end{figure}
	
\begin{figure}[t!]
  \centering \includegraphics[draft=false,keepaspectratio=true,clip,width=0.9\linewidth]{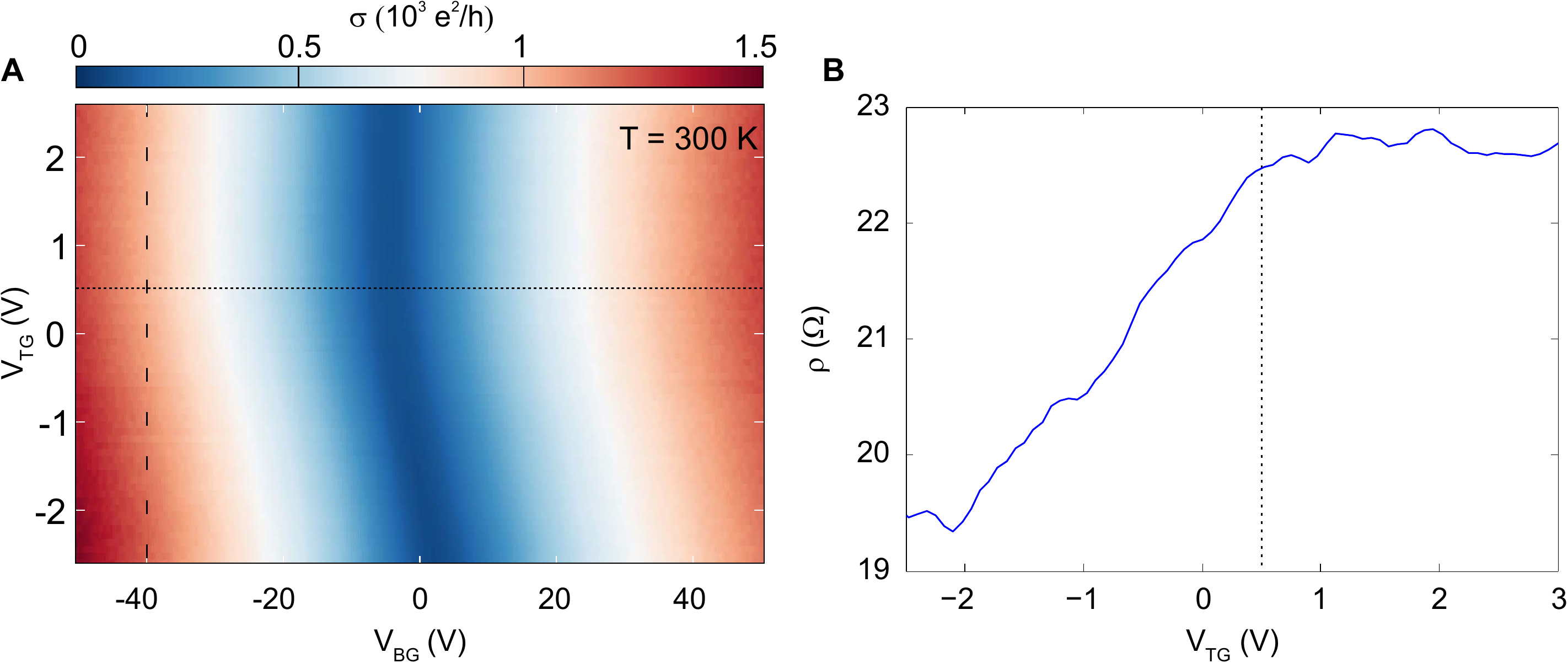}
  \caption{\small Back- and top-gate tunability of device D1. 
  \textbf{A} Longitudinal conductivity of device D1 as function of the applied top gate and back gate voltage, recorded at room temperature. 
  \textbf{B} Line cut through the data shown in panel A along $V_\mathrm{BG}=-40$~V (dashed line in panel A). The resistivity saturates as soon as carriers in the WSe$_2$ screen the top gate, indicating that graphene is screened from the top-gate by charge carriers in WSe$_2$ and that WSe$_2$ provides only a weak parallel conduction channel that is not gate-tunable.}
  \label{fS1}
  \end{figure}
 
 \begin{figure}[!htb]
    \centering
    \includegraphics[draft=false,keepaspectratio=true,clip,width=0.95\linewidth]{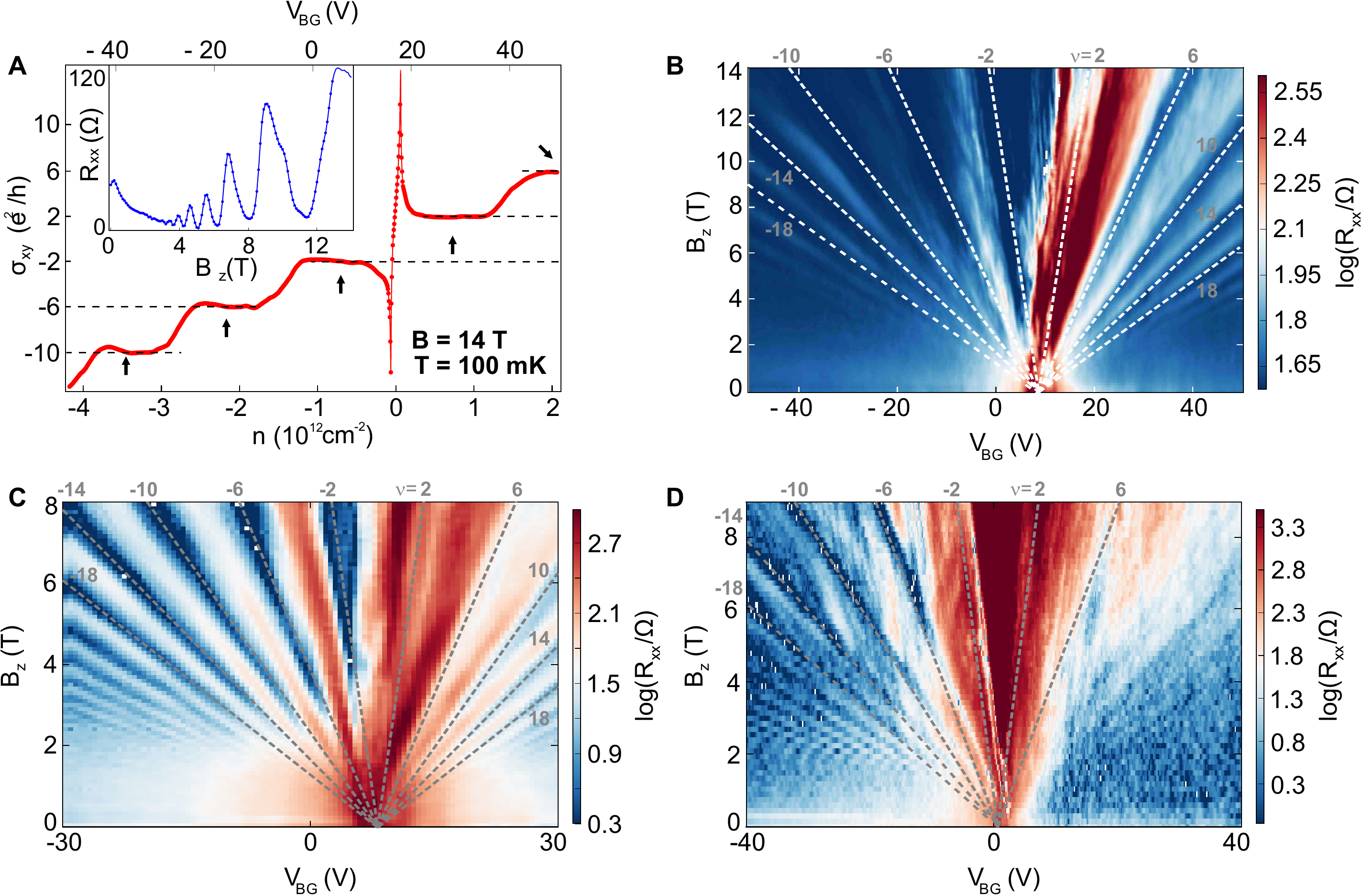}
    \caption[Fig03]{ Quantum Hall-effect measurements on WSe$_2$ devices. 
    \textbf{A} Transverse conductivity of device D1, measured in a perpendicular magnetic field of 14\,T, at 100\,mK. The half-integer quantum Hall effect and the Shubnikov-de-Haas oscillations (inset) indicate that the transport properties of the device are dominated by graphene. The overshoot around $n=0$ is caused by the presence of a weak parallel conducting channel in the WSe$_2$ layer. 
    \textbf{B} Landau fan of device D1 recorded at 100\,mK. Here and in panel~A, the voltage applied to the top gate was set to $V_{\rm TG}=-5$\,V. 
    \textbf{C-D} Landau fans of devices D2 and D3, measured at 2\,K. The dashed lines indicate the filling factors, i.e. the minima of $R_{xx}$.  Despite their high mobility, devices D1-D3 show no degeneracy lifting of the Landau levels. We attribute this fact to the local variations of the electrostatic potential seen by charge carriers in graphene caused by charge disorder in the WSe$_2$ substrate (see discussion in Sec.~\ref{sec: discussion-n*})
    }
    \label{fS2}
\end{figure}

\begin{figure}[!htb]
    \centering
    \includegraphics[draft=false,keepaspectratio=true,clip,width=\linewidth]{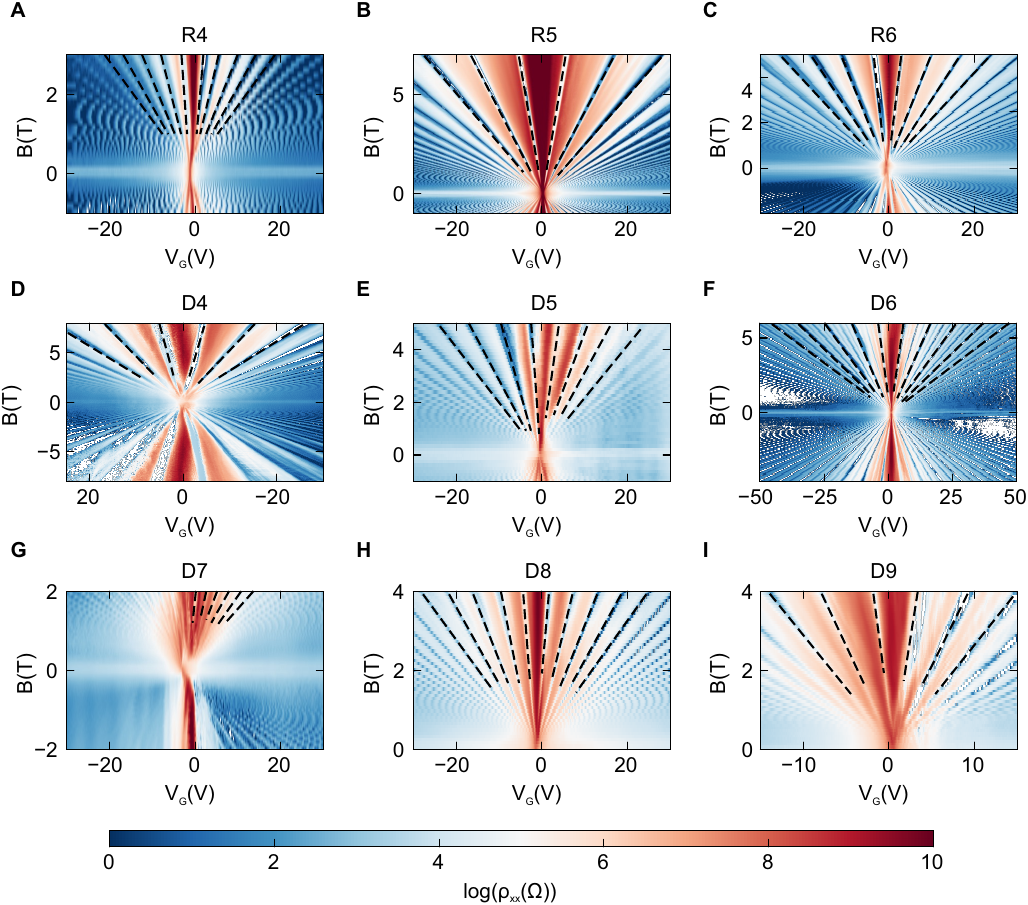}
    \caption[hall]{Quantum Hall-effect measurements on additional reference devices R4-R6 (hBN-encapsulated graphene) in \textbf{A-C} and on WSe$_2$-capped devices made from exfoliated graphene in \textbf{D-I}. The black dashed lines indicate the filling factors, i.e. the minima of $\rho_{\mathrm{xx}}$. The slopes of these lines are used to extract the lever arm for each sample (see text).}
    \label{QH_2}
    \end{figure} 

\begin{figure}[!htb]
    \centering
\includegraphics[draft=false,keepaspectratio=true,clip,width=0.8\linewidth]{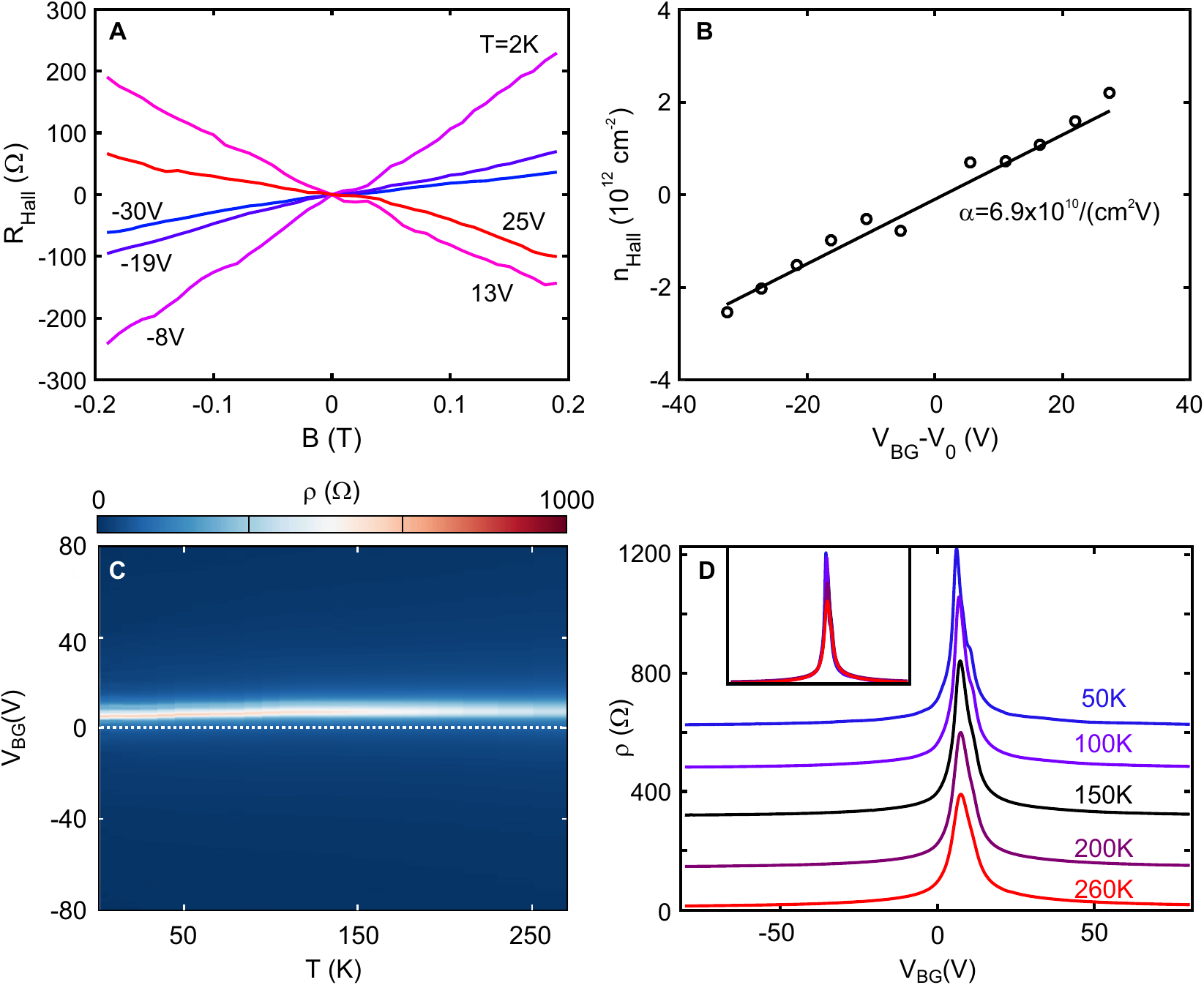}
    \caption[hall]{Standard Hall-effect measurements on device D1. 
    {\bf A} Hall resistance of device D1 at 2K, as function of the applied perpendicular magnetic field $B$. Different curves corresponds to different values of $V_{\rm BG}$. Linear fits to the data using $R_\mathrm{Hall}={B}/(n_{\rm Hall}e)$ allow extracting the charge carrier density, $n_{\rm Hall}$. 
    {\bf B} Extracted charge carrier density $n_{\rm Hall}$ as function of back-gate voltage. We use the linear fit $n_{\rm Hall}=\alpha_{\rm HE} (V_{\rm BG}- V_{\rm BG}^0)$ to extract the back-gate lever arm  $\alpha_{\rm HE}$. 
    {\bf C} 4-terminal resistivity of D1 as function of  back-gate voltages and temperature. The white dashed line indicates $V_\mathrm{BG}=0$~V. 
    {\bf D} Selected traces of C at 50~K, 100~K, 150~K, 200~K and 260~K. The charge neutrality point of device D1 remains close to $V_\mathrm{BG}=$6~V, independent of temperature, indicating that there is no charge transfer between WSe$_2$ and graphene in the whole temperature range. In these measurements, as well as in all other measurements on D1 reported in the supplement or in the main text (with the exception of those of Fig.~\ref{fS1} and Fig.~\ref{fS2}A-B) we set $V_{\rm TG}=0$~V.}
    \label{hall}
    \end{figure} 

\begin{figure}[!htb]
\centering
\includegraphics[draft=false,keepaspectratio=true,clip,width=\linewidth]{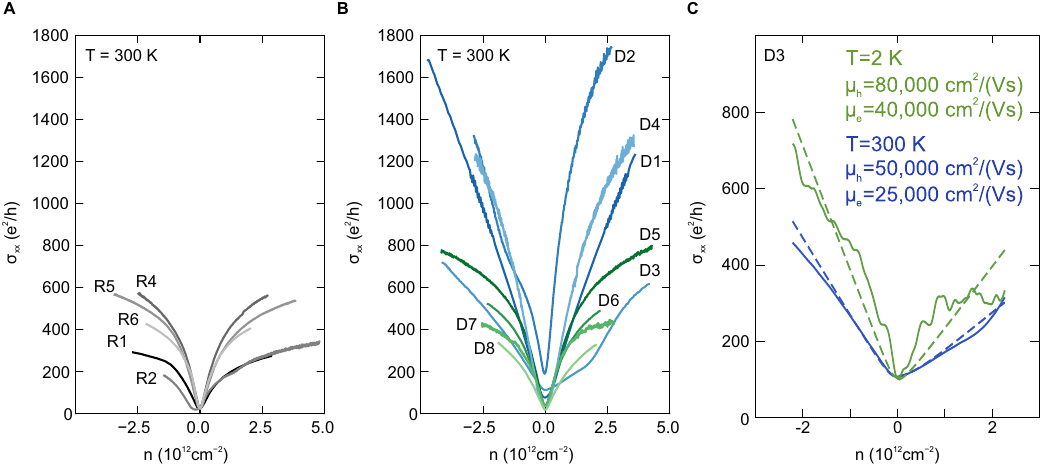}
\caption{\small Room-temperature conductivity of the measured devices. 
{\bf A} Longitudinal conductivity of the WSe$_2$/Gr/hBN devices (D1,D2,D3) and the two reference samples (hBN/Gr/hBN) (R1,R2) presented in the main text at 300~K as function of carrier density.  
{\bf B} Longitudinal conductivity of device D3 as function of carrier density  at low (2\,K) and room temperature (300 K). 
}
\label{fS3}
\end{figure}

\begin{figure}[!htb]
\centering
\includegraphics[draft=false,keepaspectratio=true,clip,width=\linewidth]{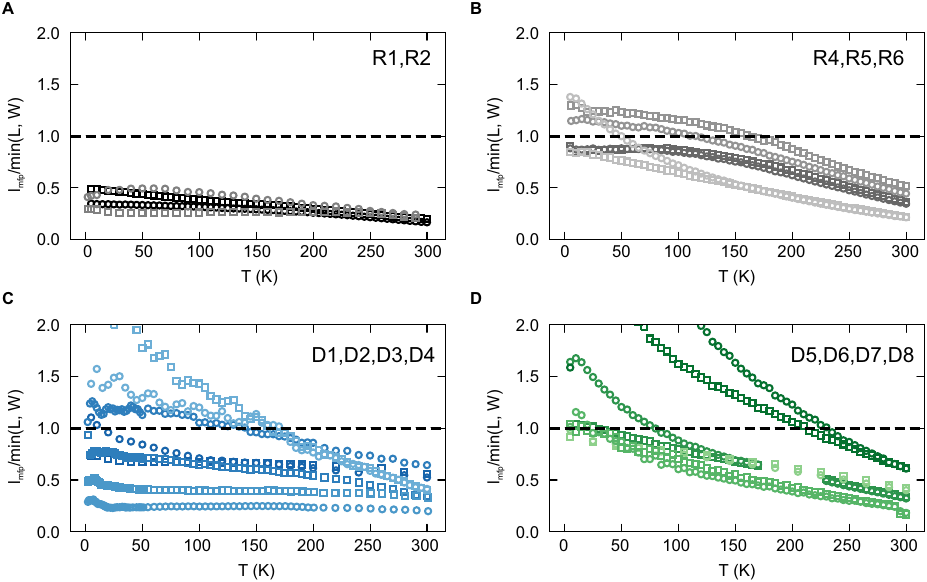}
\caption{\small Mean free path $l_{\mathrm{mfp}}$ normalized to the smallest geometric length in the device $min(L,W)$ at a density of $|n|=1.4\times 10^{12}$cm$^{-2}$ as function of temperature.
{\bf A} Devices R1-R2 {\bf B} Devices R4-R6 {\bf C} Devices D1-D4 {\bf D} Devices D5-D8.
}
\label{mean-free-path}
\end{figure}

\begin{figure}[!htb]
\centering
\includegraphics[draft=false,keepaspectratio=true,clip,width=0.3\linewidth]{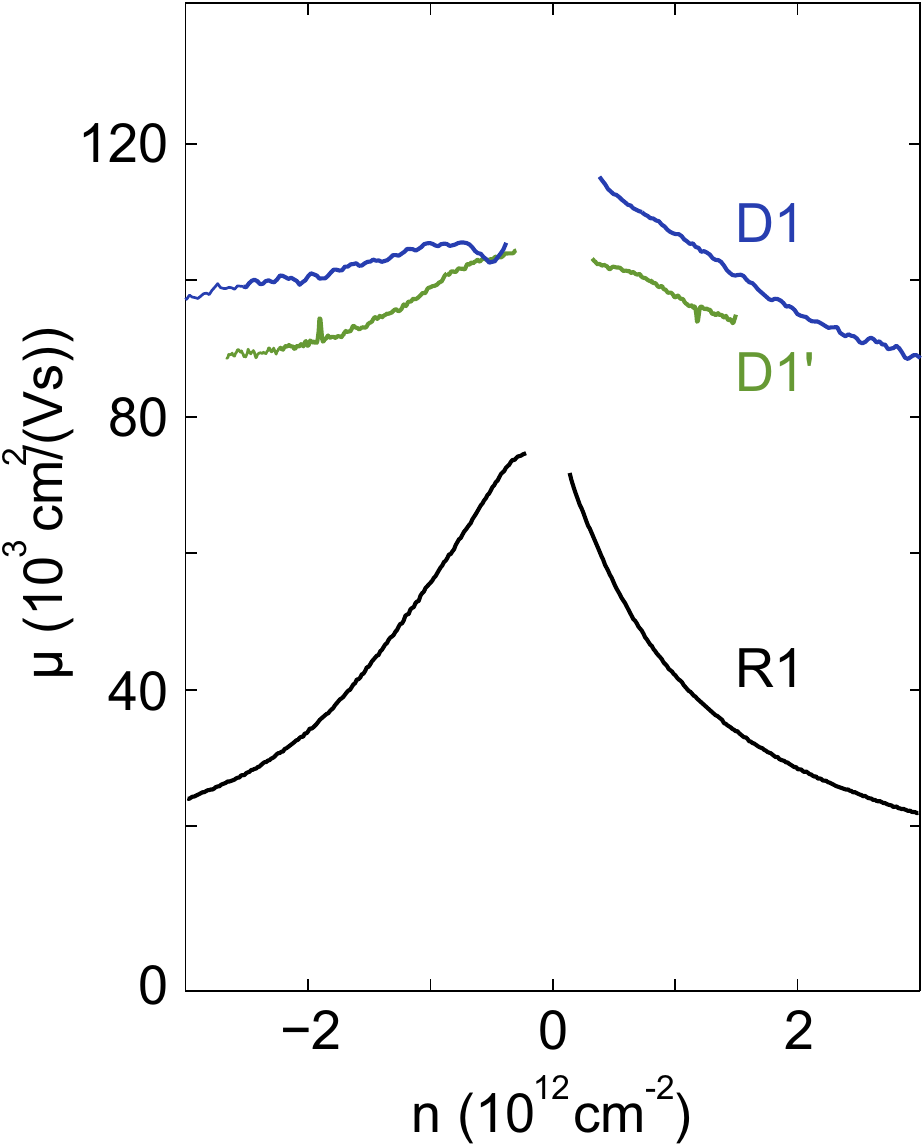}
\caption{\small Long-time stability of the WSe$_2$ devices. Room-temperature charge-carrier mobility of device D1 as function of carrier density. The traces labelled D1 and D1' correspond to two measurements on the same device (D1) taken a year apart.
}
\label{long-time-stability}
\end{figure}

\begin{figure}[!htb]
\centering
\includegraphics[draft=false,keepaspectratio=true,clip,width=0.6\linewidth]{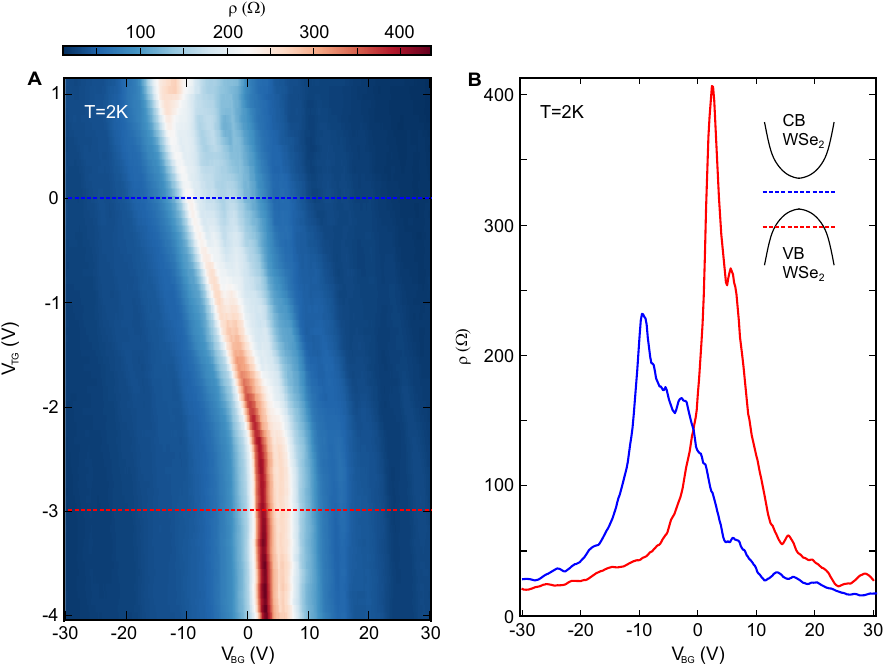}
\caption{\small  Self-screening of charge disorder in WSe$_2$.  \textbf{A} Low temperature (2~K) longitudinal resistivity of device D1  as function of the voltage applied to the top and the back gates. The resistivity peak around charge neutrality point gets lower and broader when $V_{\rm TG}$ is between -2.5 and 0.5~V, i.e. when the Fermi energy is in the band-gap of WSe$_2$.    \textbf{B} Line-cuts through the data shown in panel A along $V_\mathrm{TG}=-3$~V (red) and $V_\mathrm{TG}=0$~V (blue). The resistance peak is broad and low %and develops a considerable sub-structure 
when the Fermi energy is in the band gap of WSe$_2$, due to unscreened charge disorder in this material. } 
\label{fig:screening}
\end{figure}

\makeatletter
\let\label\SI@origlabel
\let\ref\SI@origref
\makeatother

\end{document}